# Rejection Mechanism in 2D Bounded Confidence Provides More Conformity

Sylvie Huet[1], Guillaume Deffuant[1] and Wander Jager[2]

[1] Cemagref, Laboratoire d'Ingénierie des Systèmes Complexes, France.
sylvie.huet@cemagref.fr, guillaume.deffuant@cemagref.fr
[2] Faculty of Economics, University of Groningen, The Netherlands. w.jager@bdk.rug.nl

**Abstract:** We add a rejection mechanism into a 2D bounded confidence (BC) model. The principle is that one shifts away from a close attitude of one's interlocutor, when there is a strong disagreement on the other attitude. The model shows metastable clusters, which maintain themselves through opposite influences of competitor clusters. Our analysis and first experiments support the hypothesis that the number of clusters grows linearly with the inverse of the uncertainty, whereas this growth is quadratic in the BC model.

## 1 Introduction

Understanding attitude dynamics is particularly important because attitudes motivate behaviour and exert selective effects at various stages of information processing [1]. Attitude is here understood in its psychological meaning as a tendency to evaluate a particular entity with some degree of favour or disfavour. The dynamics of attitudes are closely related to social influence, which includes individual influence on feelings, beliefs and behaviours of others [2]. These dynamics are studied, by experiments in laboratory on individuals and small groups, and are the subject of a variety of theories and assumptions. The most common assumption is a tendency of attitudes to get closer to already similar ones. A less usual assumption is a tendency to reject the other's attitude if it is psychologically uncomfortable.

This paper aims to study, through computer simulations, how individuals with both these opposite tendencies (attraction and rejection in some conditions) produce different global patterns in the attitude space. Our main result is that we observe fewer clusters than in the case of dynamics only based on an attraction. Before going through this result in more detail, we present rapidly related researches in social simulation and social psychology.

To begin, we consider the assumption of homophily. It assumes that people, especially if they are uncertain about their capacity and knowledge to evaluate a particular object, are more likely to adopt opinions and attitudes of similar others. For example, [3] shows that people like to have opinions similar to the ones of people they interact with. Similarity between receiver and source has a strong impact on the influence level of Word of Mouth [4]. Additionally, [5] suggest that homophily, facilitates the flow of information between people because of perceived ease of communication.

Secondly, besides a perspective on what drives people's attitudes towards each other, some experiments and theories focus on the forces that may drive people's attitudes apart. At the individual level, the reactance theory [6], the balance theory [7], the motivation to protect oneself [8], the social judgement [9] indicate that the persuasion can induce a rejection reaction: the behaviour, or the attitude change in the direction opposite to the persuasion effort. In groups, the social identity theory [10], the self-categorization [11] and the optimal distinction [12] theories consider a capacity to differentiate from the individuals who are members of the same group by rejecting their opinions. This rejection is usually called the "boomerang effect". The conditions of its occurrence vary from one theory to another. Furthermore, social psychologists admit that the boomerang effect remains poorly understood [13]. The social judgement theory states that uncertainty plays an important role in both attitude attraction and rejection. The social identity theory stresses that attitude rejection is linked to the salience, at a given time, of the individual social identity. At the individual level, the theories link attitude rejection to control or freedom loss, or negative relation with others. From these theories, we retain that attitude rejection occurs when several attitudes are implicitly or explicitly activated. Moreover, it is favoured by a "dissonant" situation, such as agreement on some attitudes and disagreement on others. As an example, [14] reports about students who, informed that their attitudes regarding a particular issue are close to the one of the Ku Klux Klan, decide to reinterpret this issue and to finally adopt an attitude further away from the one of the Ku Klux Klan.

Another group of interesting results for our purpose comes from the social influence paradigm which has exhibited two important group behaviours: the average consensus [15, 16] and the polarized consensus [17]. The average consensus occurs when the value of an object given by a group after discussion, is close to the average of the values given by individuals before discussion. The polarized consensus takes place when the value given by the group after discussion is significantly more extreme than the average of individual opinions before discussion. Following these studies, Nowak [18], in the social simulation domain, has recommended to investigate the tendency of individual attitudes to become more extreme (polarisation) as well as the tendency of individual to aggregate themselves in groups (clustering).

A large number of computer models are based on homophily. They postulate the existence of an attractive force between agents having close attitudes, which can be formulated using thresholds that determine when agents move towards each other's position [19-23]. This attraction threshold, also called uncertainty, can be fixed or dynamic [24, 25].

Other models, less numerous and more recent, also include a rejection mechanism in addition to the assimilation. For [26, 27] formalising the Social Judgement Theory, an individual has two thresholds on an attitude dimension: a first for assimilation and a second one for rejection (the second is assumed higher than the first). In [28], based on the theory of self-categorisation and the meta-contrast principle, an individual tends to minimise the distance to a prototypical opinion which defines his own group and, at the same time, he maximises the distance to an external group. Moreover, a rejection effect appears in [29, 30] as an emerging effect of homophilic individual interactions. This effect is due to the fact that getting closer in the 2-dimension space of attitude may in some cases result in a shift away on the global attitude (which is a weighted aggregation of the attitudes).

The attitude dynamic model we propose postulates multidimensional attitudes, like in [27, 29-36]. Considering two dimensional attitudes, our main assumption is that, if you strongly disagree with someone on an attitude-dimension 1, and are close on a attitude-dimension 2, you tend to solve the dissonance by shifting away on the attitude-dimension 2. More precisely, when attitudes on both dimensions are far or close from each other, we follow the hypotheses of bounded confidence (BC) models [19, 21, 23-25]: when both are close, the attitudes tend to get closer, when both are far, there is no influence. Therefore, our model is similar to a multi-dimensional bounded confidence model, except that we added the rejection attitude when people are close on one attitude and far on the other.

The following part of this paper describes the model in the ODD framework [37]. Then, we present examples of simulations for different parameters, which lead to hypothesise that the number of clusters grows linearly with the inverse of the uncertainty. Then, we show some results of more systematic exploration of the parameter space which support this hypothesis. We finally discuss the results and conclude.

## 2 Overview of the model

### 2.1 Purpose of the model

The purpose of the model is to test the collective effects of a particular rejection dynamics in 2-dimensional bounded confidence models which are based on a individual attraction dynamics. The rejection takes place when individuals are close on one attitude and far on the other.

### 2.2 State variables and scales

We consider a population of $N$ individuals, each having a 2-dimensional attitude or two different attitudes $x_1$ and $x_2$, represented by real numbers between -1 and +1, and the related uncertainties $u_1$ and $u_2$. Uncertainty is a term used for convenience, because this variable may represent conviction and openness to others as well. It corresponds also to the latitude of acceptance of the Social Judgement Theory and represents the level of ego-involvement in the value of the attitude. In the following experiments all individuals have the same uncertainties, on both attitudes.

### 2.3 Process overview and scheduling

At each time step, we choose a pair of individuals $A$ and $B$ at random, and they may influence each other. More precisely, at each time step, the algorithm is as follows:

```
N/2 times repeat:
    • choose couple of individuals (A,B) at random;
    • A may influence B,
      B may influence A.
```

The influence depends on conditions on the values of attitudes and uncertainties. Suppose $A$ has attitudes $x_1$ and $x_2$ with uncertainties $u_1$ and $u_2$, and $B$ has attitudes $x_1'$ and $x_2'$ with uncertainties $u_1'$ and $u_2'$. Then, $A$ compares its attitudes with the ones of $B$. Three cases arise:

- Case 1: $B$ is close to $A$ on both attitudes:

$$|x_1 - x_1'| \leq u_1 \quad \text{and} \quad |x_2 - x_2'| \leq u_2 \tag{1}$$

Then both attitudes of $A$ get closer to the ones of $B$:

$$x_1 := x_1 + \mu.(x_1' - x_1)$$
$$x_2 := x_2 + \mu.(x_2' - x_2) \tag{2}$$

Where $\mu$ is a kinetic parameter of the model. This parameter represents the velocity of the attraction or the rejection. In our following study, $\mu$ has the same value for all individuals.

- Case 2: $B$ is far from $A$ on both attitudes:

$$|x_1 - x_1'| > u_1 \quad \text{and} \quad |x_2 - x_2'| > u_2 \tag{3}$$

Then, there is no influence of $B$ on $A$.

- Case 3: $B$ is far from $A$ on one attitude and close to $A$ on the other. Without loss of generality, we suppose:

$$|x_1 - x_1'| \leq u_1 \quad \text{and} \quad |x_2 - x_2'| > u_2 \tag{4}$$

Then two cases arise, depending on whether $A$ and $B$ differ strongly on attitude 2. We introduce the positive parameter $\delta$, ruling the intolerance threshold which globally depends on the uncertainty, i.e. on the ego-involvement level:

  o Case 3.1: $A$ and $B$ do not differ strongly on attitude 2

$$|x_2 - x_2'| \leq (1+\delta).u_2 \tag{5}$$

Then, the disagreement is not strong enough to trigger the rejection. $A$ approaches $B$ on attitude 1 and ignores $B$ on attitude 2:

$$x_1 := x_1 + \mu.(x_1' - x_1) \tag{6}$$

  o Case 3.2: $A$ and $B$ differ strongly on attitude 2

$$|x_2 - x_2'| > (1+\delta).u_2 \tag{7}$$

Then, *A* shifts away from *B* on attitude 1. The movement is proportional to the distance needed to get $x_1'$ out of *A*'s range of uncertainty around $x_1$.

$$x_1 := x_1 - \mu.sign(x'_1 - x_1).(u_1 - |x_1' - x_1|) \qquad (8)$$

Where *sign*() is the sign function, which returns -1 if its argument is strictly negative, +1 otherwise. Moreover, we confine the attitude within the bounds (-1, +1) of the attitude space:

$$\text{If } |x_1| > 1 \text{ then } x_1 := sign(x_1) \qquad (9)$$

The following figures illustrate the different types of interactions (attraction, rejection or indifference). Note that we suppose that uncertainties are the same on both dimensions and for all individuals. This means that we only get symmetrical interactions: if *A* attracts *B*, *B* attracts *A*; if *A* rejects *B*, *B* rejects *A*.

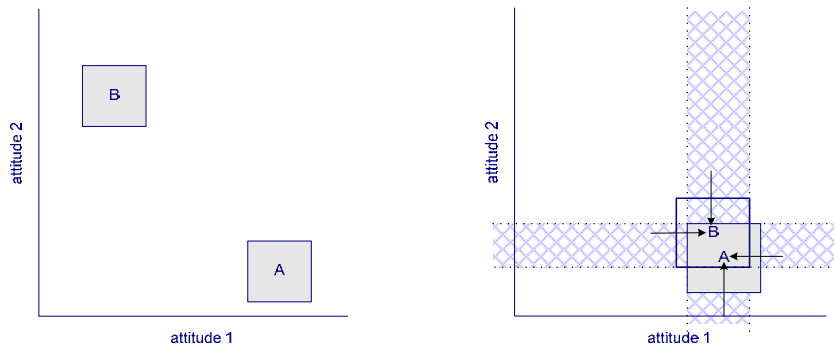

**Fig. 1.** *A* and *B* in a situation of no influence on both dimensions (left) and in situation of attraction (right)

Figure 1 shows on the left the case where *A* is not influenced by *B*: they are far from each other on both dimensions. On the right, figure 1 shows the case where *A* is attracted by *B* and vice-versa because they are close to each other. This means each one has his attitude in the other's acceptance zone.

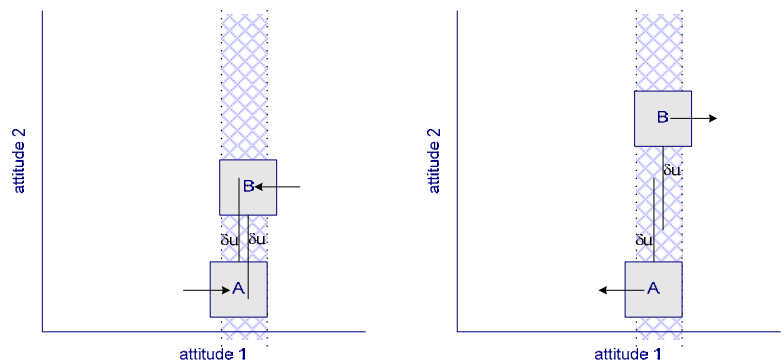

**Fig. 2.** Left: *A* and *B* in a situation of attraction on one dimension (on attitude 1 dimension here) and indifference on the other dimension. Right: *A* and *B* in a situation of rejection on one dimension (on attitude 1 dimension here) and indifference on the other dimension

Figure 2 left shows another case where people are close to each other on only one dimension. People are far from each other on one dimension but not enough far to consider the proximity on the other dimension as unacceptable. Thus, they assimilate each other on the dimension where they are close. On the contrary, figure 2 right shows the cases where people are far enough from each other on one dimension. The proximity on the other dimension is perceived as unacceptable. Thus, they move away from each other on this dimension.

### 2.4 Initialisation

We consider a population of 1000 individuals with two attitudes. On each dimension, the attitude is randomly initialised following a uniform distribution comprised between -1 and +1. Uncertainty $U$ is constant and identical on each dimension. Identically, $\mu$ has the same value for all individuals.

## 3 Analysis of several examples

In this section, we observe several simulation examples, and this analysis leads to formulate the hypothesis that the number of clusters is a linear function of $1/U$.

### 3.1 Evolution with uncertainty $U = 0.2$ and intolerance threshold with $\delta = 0$

Figure 3 shows an example of evolution for uncertainties $U = 0.2$, and intolerance parameter $\delta = 0$, and the kinetic parameter $\mu = 0.3$. The number of time steps $t$ appears on the top of each picture.

On Fig. 3, both attitude axes are represented; black spots indicate the attitude position of individual agents. On this figure, we observe that the population is progressively organised into several clusters. The clusters are not regularly organised on horizontal and vertical lines, as observed with the classical bounded confidence model. They tend rather to be located on oblique lines, which are not strictly regular.

Moreover, the individuals oscillate in the clusters, with constant amplitude of oscillation, leading to a permanent diversity within the cluster. The reason is that individuals are pushed away from the cluster by other clusters, located close on one dimension and far enough on the other. These movements compensate each other because generally there are several neighbouring clusters that reject the cluster in opposite directions. Moreover, individuals are attracted by the cluster itself, especially if the cluster includes many individuals. Therefore one can say that the clusters are metastable, because if there is a strong perturbation (deletion of a neighbouring cluster) this may dramatically modify the equilibrium. This is a big difference with the classical bounded confidence model, in which after a while, clusters keep concentrating with time, each independently from the others.

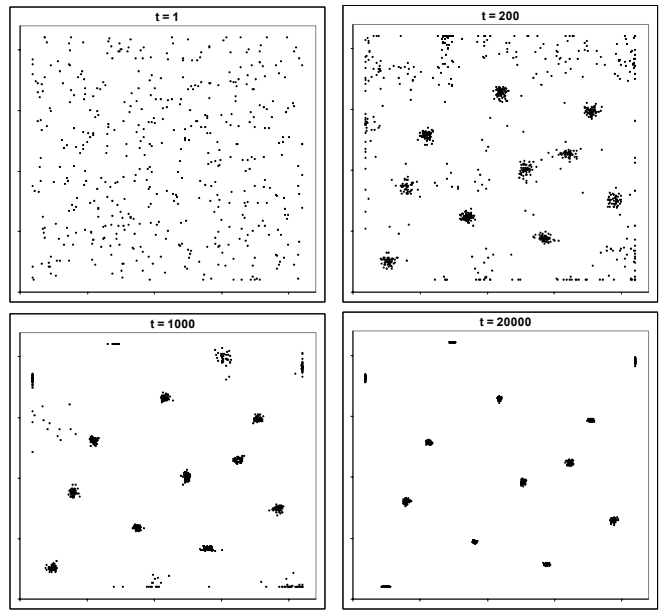

**Fig. 3.** Initial population uniformly distributed in 2D attitude space. $U = 0.2$, $\mu = 0.3$, $\delta = 0$. We observe the emergence of metastable clusters, with remaining oscillations of individuals within the clusters. Moreover, some flat clusters are located on the borders of the attitude domain, containing radicalised individuals. See text for discussion.

A second important difference with the classical BC model is that we get clusters on the border of the attitude domain. With the BC model, the first clusters are always inside the attitude domain, of a distance which is about the uncertainty $U$. On the contrary, with this model, it seems that there are always some clusters which have a larger attitude than any individual at the initialisation. These border clusters are flat, because their neighbouring clusters tend to push them away outside the attitude domain. This is a polarisation phenomenon in the sense of Nowak: a part of individuals gets more extreme. If we removed the constraint to remain within the bounds of the attitudes, the global range of attitude would grow, and we would finally end up with stable clusters, not disturbing each other, in a significantly larger attitude domain.

### 3.2 Spatial organisation of the clusters and hypothesis of linearity of their number with 1/U

The spatial organisation of the clusters can be further analysed. In this particular case where $\delta = 0$, we note that there is only one cluster on a horizontal or vertical line. Indeed, two clusters on the same horizontal or vertical line is an unstable situation. If the clusters are far, they tend to push each other from the line. If the clusters are close, they tend to merge. This can be checked by considering the histogram of presence of the individuals on each axis on fig. 4. We note that 13 clusters appear on the projection of both axes. Moreover, the distance between the clusters is too small to prevent the rejection to play (11 clusters is the maximum, to provide a distance of at least $U$ between two consecutive clusters), which explain why the individuals oscillate in the clusters.

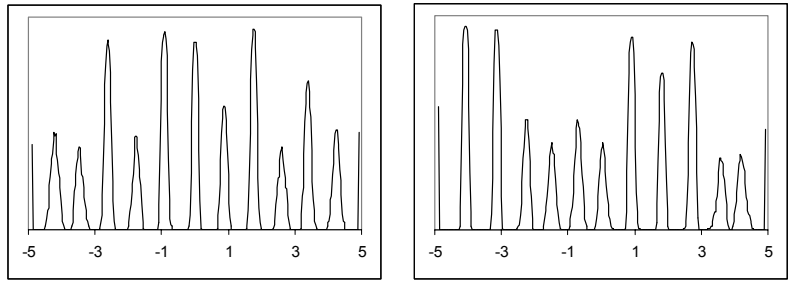

**Fig. 4.** Kernel density estimator on horizontal axis (left) and vertical axis (right), for the final situation of fig. 4 ($U = 0.2$, t = 20000). One notes that the 13 final clusters are regularly distributed on each axis.

In this case, the number of clusters can be analysed on a single axis: there should be a minimum interval between the clusters on each axis which is about the value of $U$. As we have seen, because of the metastability, it is possible to get slightly smaller intervals. Nevertheless, one can expect a number of clusters varying linearly with $1/U$.

### 3.3 Influence of intolerance threshold $\delta > 0$

When the intolerance threshold gets higher, the conditions for rejection are more restricted: the disagreement on one attitude must be higher. Figure 5 shows two examples of final attractors, for $U = 0.2$, $\delta = 1$ (left) and $\delta = 1.5$ (right). The number of clusters seems to increase with $\delta$.

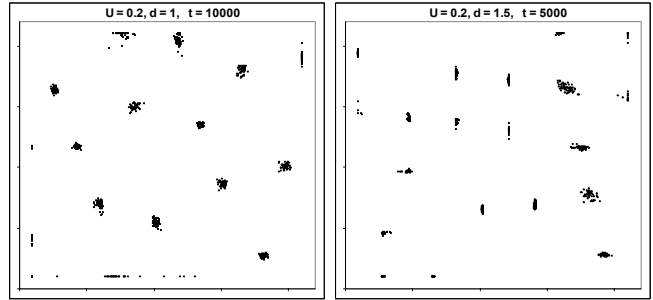

**Fig. 5.** Example of final configurations for $U = 0.2$, $\mu = 0.3$, $N = 1000$. $\delta = 1$ (left), $\delta = 1.5$ (right). It is possible to get 2 clusters on the same horizontal or vertical line, which is unstable when $\delta = 0$. Moreover, for $\delta = 1.5$, some clusters are flat inside the attitude domain.

We observe that for these values of $\delta$, it becomes possible to get two clusters on the same horizontal or vertical line, when they are not too far apart (they remain in the tolerance zone). This explains why there are more clusters. Nevertheless, we can hypothesise that this number should still vary linearly with $1/U$, but with a higher coefficient.

Moreover, for $\delta = 1.5$, we observe flat clusters inside the attitude domain, whereas this did not take place for $\delta = 1$. Such a flat cluster appears when all the neighbour clusters are on the same line in the tolerance zone, or far on both attitudes. The rejection interactions are therefore only in one direction.

### 3.4 Different values of uncertainty $U$ with intolerance threshold $\delta = 0$

Figure 6 shows several attractor configurations for different values of uncertainty $U$. This first exploration suggests that the number of clusters decreases with $U$, like with the BC model. The observations made on our first simulation extend to these cases: Oscillations of individuals remain, with higher oscillations when $U$ increases, and spatially organised to avoid two clusters on the same horizontal or vertical line. In each case, we get flat clusters with the maximum value for one attitude (polarisation).

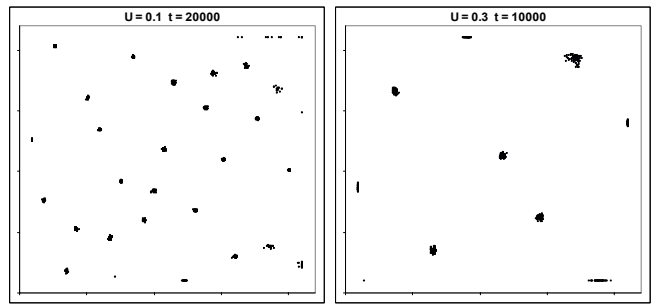

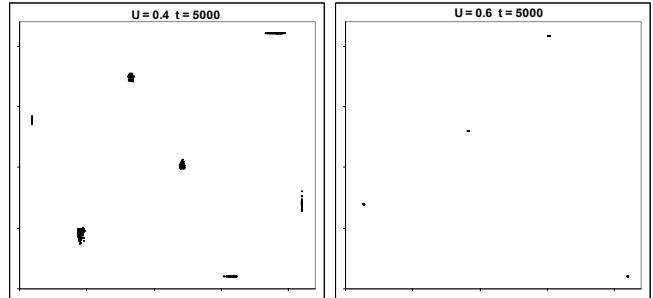

**Fig. 6.** Examples of attractor configurations for different values of uncertainty $U$ and intolerance parameter $\delta = 0$, $\mu = 0.3$. Population size $N = 1000$.

For $U = 0.6$, we observe that the clusters become very concentrated, like in the simple BC model. The reason is that with 4 clusters, the intervals between the clusters on a same horizontal or vertical line can easily be higher than $U$, and therefore avoid to generate a competition between the attraction in the cluster and the rejection from the neighbour clusters.

## 4   Systematic analysis of the number of clusters

We are interested in comparing the final number of attitude clusters with the one given by standard BC model.

### 4.1   Computing the number of clusters

From the individual-based simulations, we collect the average, minimum and maximum final number of clusters. To compute the number of clusters, we define a minimum distance $\varepsilon$ between attitudes, below which we consider that they belong to the same cluster. We compute the clusters as groups of agents such that between any couple of agents of opinions $x$ and $x'$ in the group, there is a list of agents in the group of opinions $(x_1, x_2, …, x_k)$ making a chain of couples with an Euclidian distance below $\varepsilon$. The following pseudo-code can be used to compute the clusters; `necessaryToLookAt` is a table containing the identification number of each individual for all the population:

```
for all i of the population
 if necessaryToLookAt[i] > 0
   currentCluster.add(i)
   compt++;
   necessaryToLookAt[i] = 0
   while currentCluster.isNotEmpty()
     for all j of the population
       if necessaryToLookAt[j] > 0
         if distance(pop[currentCluster.get(0)],pop[j]<epsilon)
           necessaryToLookAt[j] = 0
           currentCluster.add(j)
```

```
        compt++
    currentCluster.remove(0)
  nbClusters++
 if compt = populationSize then i = populationSize
```

In practice, we chose $\varepsilon = 0.2\ U$, and we neglected the clusters of size lower or equal to 3 individuals. The simulations are stopped after 1,000,000 iterations. They can be stopped before if the number of clusters has not changed after 100,000 iterations.

### 4.2 Final number of clusters

The BC model, in one dimension, yields a final number of clusters $n_c$ in a population initialised with a uniform law on an attitude space of width $2M$, with all the same uncertainty $U$, which can be approximated by:

$$n_c \approx \frac{M}{U} \tag{10}$$

In the 2-dimensional case, when both attitude axes are adjusted independently and all have the same uncertainty $U$ on both attitude dimensions, this rule is repeated on all lines of the space, therefore we get:

$$n_c \approx \left(\frac{M}{U}\right)^2 \tag{11}$$

This result is confirmed by figure 7 which presents on abscissa $1/U^2$ and on y-axis, the average number of clusters obtained on 30 replicas.

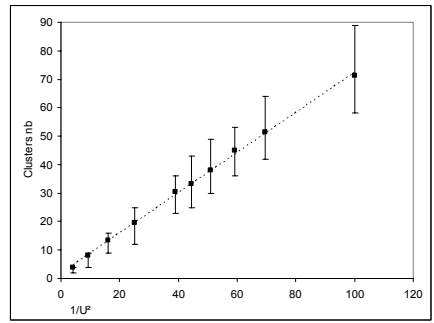

**Fig. 7.** Average final number of clusters of the 2D bounded confidence model as a function of $1/U^2$. Error bars indicates minimum and maximum obtained on 30 replicas.

Figure 8 shows the number of clusters obtained with rejection dynamics, for different values of $U$ and $\delta$. These results confirm the hypothesis of linearity of the number of clusters with $1/U$ for $\delta = 0$ and $\delta = 0.5$ (left).

For $\delta = 1, 1.5, 2$ and $3$, there is a non-linearity for $U$ larger than 1 (only 1 and 2 are presented on the figure). When $U$ is larger than 0.3, and $\delta$ is large, the conditions for rejection are much constrained by the size of the domain: two individuals must be at both sides of the domain. Most of the interactions correspond therefore to the standard BC, and the curve is therefore quadratic. When $U$ decreases ($1/U$ grows), the rejection becomes more common and the curve becomes linear.

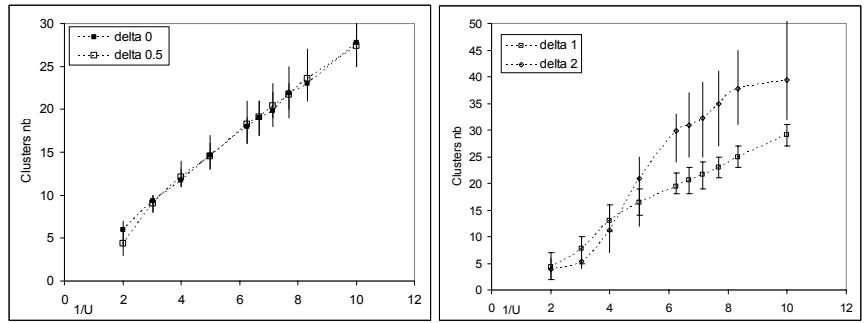

**Fig. 8.** Mean final number of clusters for the model with rejection as a function of $1/U$, for various values of $\delta$. $N = 1000$ and $\mu = 0.3$. The error bars are the minimum and maximum numbers met in 30 replicas. On the left, for $\delta = 0$ and $\delta = 0.5$, the number of clusters seems linear with $1/U$. On the right, the behaviour is not linear for large $U$.

## 5 Discussion, conclusion

In the model of 2-dimensional attitude dynamics we propose, one shifts away from a close attitude on one axis when the interlocutor is far on the other axis. We assume that this is a way to solve the dissonance between the attitude axes. The distance threshold to trigger rejection depends on the intolerance parameter $\delta$ and on the uncertainty $U$, which may define a non-commitment zone, in which the dissonance is tolerated. When the conditions of rejection are not met, the model behaves exactly like the 2D bounded confidence (BC) model.

The first explorations of this model, in the simple case where all uncertainties are the same, show several striking results, in comparison with the 2D BC model:

- The dynamics generally leads to several clusters, which are in competition and tend to reject each other. Finally, the system reaches a metastable state: the stability is due to contradictory rejections from neighbouring clusters, which compensate each other. If one of its neighbouring clusters is removed, the position of a cluster changes significantly, and it may even disappear. Moreover, individuals belonging to a cluster are in constant movement, with amplitudes depending on the cluster size and on the proximity of competing clusters. In this respect, the configuration is very different from the one obtained with simple BC model where, after a while, clusters keep concentrating with time, each independently from the other.

- Several clusters are often smashed on the limits of the attitude domain. This may be interpreted as a radicalisation of a part of the population, which reaches the maximum absolute value of one of the attitudes. This never happens with the 2D BC model.

- In the case where the intolerance threshold $\delta = 0$, two clusters cannot be maintained on the same horizontal or vertical line. Therefore, the clusters tend to occupy points of the space where they are as far as possible from other clusters on each axis. This analysis suggests a number of clusters growing linearly with $1/U$, whereas the cluster number grows quadratically with $1/U$ in the 2D BC model. When $\delta$ grows, configurations with more than one cluster on a line may be stabilised, but this number is limited by the size of the tolerance zone. Therefore, the growth of the cluster number should still be linear, but with a factor growing with $\delta$. First systematic experiments support this statement.

These results suggest several points to discuss.

The metastability of the clusters is due to the bound we impose on the attitude. Indeed, without this bound, the attitudes grow until the distance between the clusters is higher than the uncertainty in all directions. Then, the clusters do not influence each other, and they keep concentrating as in the BC model. It is probable that the final number of clusters is close to the one obtained with the bound on attitude, but we did not check this systematically. In any case, the metastability of the clusters is an interesting feature of this model, which fits better real groups than the perfect similarity obtained without a bound (or by a standard BC model).

Even without bounds, we obtain a global result which shows strong similarities with social identity and self-categorization theories. Our individuals tend to minimise their in-group distance and maximise their out-group distance (to competing groups). We also get some polarized groups (which have more extreme opinions than all the individuals initially). This reminds the results of Moscovici and Doise. Therefore, with a model considering only pair interactions, we get group dynamics which seem to make sense in a social psychology perspective.

However, the model remains very rough, and one should check if these interesting properties remain when adding more sophisticated hypotheses. In particular, in our model, all attitudes are considered to have the same weight on the behaviour, whereas one expects that only disagreements on attitudes deeply related to social identity can lead to rejection. To take this aspect into account, we thus should consider attitudes different of different types.

In the future, we also intend to continue to explore the properties of this model. In particular, we suspect interesting effects of the population size on the number of clusters. Furthermore, introducing extremists like in [24] could produce unexpected effects.